\documentclass[
twocolumn,
superscriptaddress,
 amsmath,amssymb,
pra,
]{revtex4-2}
\usepackage{upgreek}
\usepackage{graphicx}
\usepackage{dcolumn}
\usepackage{bm}
\usepackage{xcolor}

\usepackage[colorinlistoftodos,prependcaption,textsize=tiny]{todonotes}

\newcommand{\operator}[1]{\widehat {\mathrm {#1}}}
\newcommand{\bra}[1]{\mbox{$\langle#1|$}}
\newcommand{\ket}[1]{\mbox{$|#1\rangle$}}

\begin{document}

\preprint{APS/123-QED}

\title{Efficient lambda-enhanced gray molasses using an EIT-based laser locking scheme }

\author{Timothy Leese}
\affiliation{%
 School of Physical Sciences, Faculty of STEM, The Open University, Walton Hall, Milton Keynes, MK7 6AA, UK
}
\affiliation{%
 Department of Physics, University of Oxford, Parks Road, Oxford,
OX1 3PU, UK
}

\author{Siobhan Patrick}
\author{Silvia Bergamini}
\author{Calum MacCormick}
\affiliation{%
 School of Physical Sciences, Faculty of STEM, The Open University, Walton Hall, Milton Keynes, MK7 6AA, UK
}
\email{c.maccormick@open.ac.uk}

\date{\today}

\begin{abstract}
We present a novel implementation  of  lambda-enhanced gray molasses cooling in  a non-standard beam geometry and with an inexpensive laser locking set-up that only provides limited coherence of the Raman laser beams.  In contrast to the established use of resource-intensive  phase locking methods, our laser system uses two independent lasers, frequency -locked to a spectral feature produced by an  electromagnetically induced transparency (EIT) resonance. We show that this approach achieves sufficient coherence to enable effective gray molasses cooling without the need for costly GHz electronics, significantly reducing the complexity and cost of experimental setups and represents a step toward more accessible cold atom technologies. Furthermore, the cooling remains efficient even with a non-optimal beam geometry, typical of cold-atoms quantum computing platforms based on optical tweezers. A  wave-function Monte Carlo analysis supports the experimental findings, offering insight into the cooling dynamics of this unconventional scheme. 
\end{abstract}           
\maketitle

Laser cooling methods are widely used in quantum computing and precision measurements. Often, experiments that rely on laser cooling may benefit if the final temperature is lower; as an example, evaporative cooling methods common in experiments studying Bose or Fermi gases may be more efficient when the phase space density is maximised prior to the evaporation (see the discussion in \cite{rosi2018}). Quantum computing with neutral atoms has gained significant traction due to its scalability, long coherence times, and the ability to implement high-fidelity quantum gates using Rydberg interactions \cite{PhysRevLett.129.200501}\cite{PhysRevLett.104.010503}\cite{PhysRevLett.104.010502}\cite{PhysRevX.12.021049}. Strong dipole-dipole interactions between highly excited Rydberg states enable fast and deterministic entangling gates, a critical component for universal quantum computation \cite{PhysRevX.12.021049} and applications such as quantum sensing~\cite{MacCormick2016}, or non-universal models of quantum computing~\cite{Mansell2014}. However, the fidelity of these gates is highly sensitive to atomic motion, thus making temperature control a crucial factor in optimising performance.
Advances in cooling and optical trapping have enabled neutral atom arrays with well-controlled spatial positioning, significantly improving coherence times and interaction strengths \cite{PhysRevLett.129.200501}, \cite{PhysRevLett.104.010503},\cite{PhysRevLett.104.010502}, but the fidelity is still limited by the localisation of the atoms in the arrays.
It is therefore necessary to implement strategies for cooling down to a few microkelvin. One effective approach is the use of lambda enhanced gray molasses cooling, which has been demonstrated to achieve $\sim 4 -20 \,\upmu\text{K}$ temperatures in various atomic species \cite{rosi2018}~\cite{PhysRevA.87.063411}~\cite{Fernandes2012}.

In this work, we show that using simple and inexpensive method of laser locking two diode lasers using MHz frequency electronics can implement lambda-enhanced gray molasses cooling as effectively as much more complicated and expensive techniques based on GHz frequencies. 
Unlike conventional methods that require stringent phase-locking of the cooling lasers~\cite{agnew2024}, our approach remains effective even when the laser fields are not phase-locked, making it more practical for experimental implementation. Furthermore, we test the efficiency of the cooling using a non-ideal beam geometry that is typically required due to constrained optical access in complex quantum computing platforms based on optical tweezers. We achieve efficient cooling that we calculate (based on a simple model of an electromagnetic induced transparency (EIT) based CNOT gate) enhances Rydberg gate fidelity from $1-F=0.029$ at $T=65 \, \upmu\text{K}$ to $1-F=3.1\times10^{-4}$ at $T=6.8 \, \upmu\text{K}$, while minimising additional experimental complexity.
We present a detailed analysis of this method, including numerical simulations and experimental considerations. Preliminary experimental results demonstrate the effectiveness of this method in reducing the atomic sample's temperature by one order of magnitude and suppressing atomic motion. Our approach provides a scalable and cost-effective solution for improving neutral atom quantum computing fidelity and quantum sensing.
One further aspect of our experiment is noteworthy. The geometry of the cooling beams has been compromised to accommodate high numerical aperture lenses, which enable very tight optical dipole trapping with a far red-detuned laser beam, and simultaneously, very high-resolution atomic detection via fluorescence imaging. In spite of this experimental arrangement, the gray molasses is found to be very efficient.

Our paper is organised as follows. We begin by discussing the EIT-offset locking scheme, which is at the heart of our technique and which provides for well stabilized lasers capable of preparing a coherent dark state. We then discuss the application of these lasers to gray molasses, where we present our results. Finally, we compare our results to a wave function Monte Carlo simulation of our experiment. 
\section{Experimental setup}
\subsection{EIT-based offset lock}
All our experiments study $^{87}\text{Rb}$. The light for our gray molasses experiments is produced using a laser system built from two commercial diode lasers. The coupling laser (a TOPTICA DL Pro) is stabilised to the $5S_{1/2},F=2\to 5P_{3/2}, F^{\prime}=1,\,3 $ saturated absorption crossover peak, using TOPTICA's Digilock module. The Digilock modulates the laser current at 40 kHz and we use the Digilock's automated routines to optimise the laser stability.  The Raman laser, (a TOPTICA DL 100) is locked to an EIT feature produced by co-propagating both the coupling laser and Raman laser through an Rb vapour cell. Thus, the Raman laser is locked to the coupling laser using a method similar to that used in \cite{Abel2009},~\cite{bell2007}. The use of a narrow EIT feature reduces the \emph{relative} frequency noise, and we estimate the two lasers are locked so that their beat-note frequency linewidth (\lq Raman' linewidth, denoted $\Gamma_\text{R}$) is at most $\Gamma_\text{R}\sim10\,\text{KHz}$), similar to that demonstrated in \cite{bell2007}.

The Raman laser is coarsely tuned near the $F=1\to F^{\prime}=1,\,3 $ hyperfine transition by adjusting the laser diode current, temperature and the voltage applied to the PZT, which controls the diode laser's external cavity length. A `probe beam' is picked off the Raman laser output and passed through an EOM, modulated at 10 MHz with a modulation depth of 0.03, to generate sidebands.

\begin{figure}
    \centering
    \includegraphics[width=0.9\linewidth]{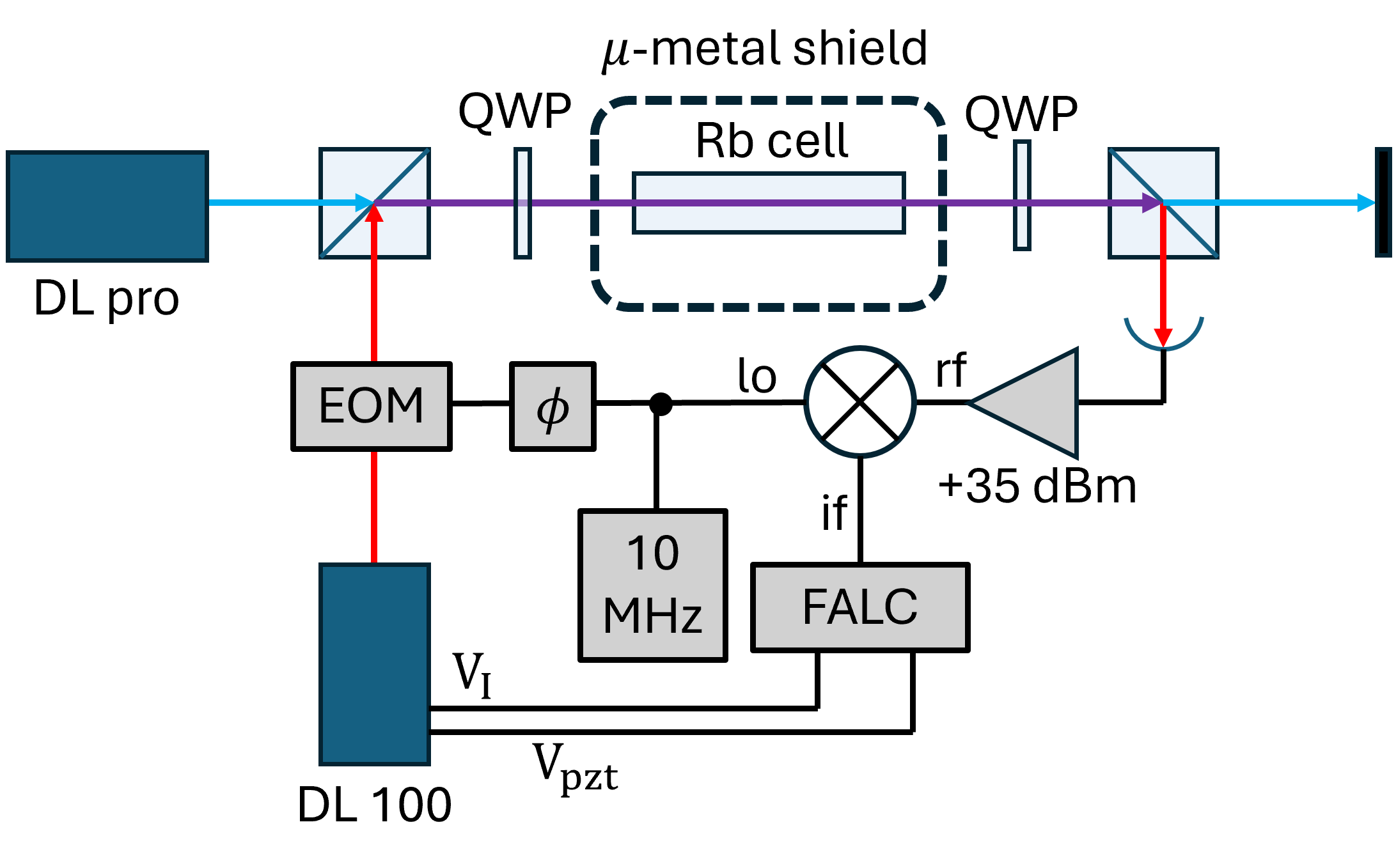}
    \caption{The EIT offset locking scheme. The DL Pro laser is stabilised to the $F=2\to F^\prime = 1,3$ saturated absorption crossover peak (the saturated absorption spectrometer is omitted for clarity). The DL 100 laser is phase modulated with a 10 MHz EOM, and then mixed on a polarizing beamsplitter cube with the stabilised DL Pro laser. Both are then passed through a quarter waveplate (QWP) and the rubidium vapour cell, located in a $\upmu-$metal housing. After propagating through the cell, the DL100 beam is picked off using a polarising beam splitter cube and sent to a fast photodiode. The fast photo diode output is mixed with the 10 MHz local oscillator signal and amplified by the TOPTICA FALC module. }
    \label{fig:enter-label}
\end{figure}

A portion of the coupling laser beam—stabilized as described above—is also picked off and spatially overlapped with the modulated probe beam using a polarizing beam splitter. The collimated, overlapped beams, have a waist of $\sim 1.5\,\text{mm}$, are then passed through an $\times 4$ beam expander, which reduces their intensity to minimize power broadening of the EIT features while preserving signal strength. 
Following the beam expander, a quarter-wave plate circularly polarizes the two beams with opposite handedness. The beams then co-propagate through a $10\,\text{cm}$ rubidium vapor cell.
\par
\begin{figure}
    \centering
    \includegraphics[width=1\linewidth]{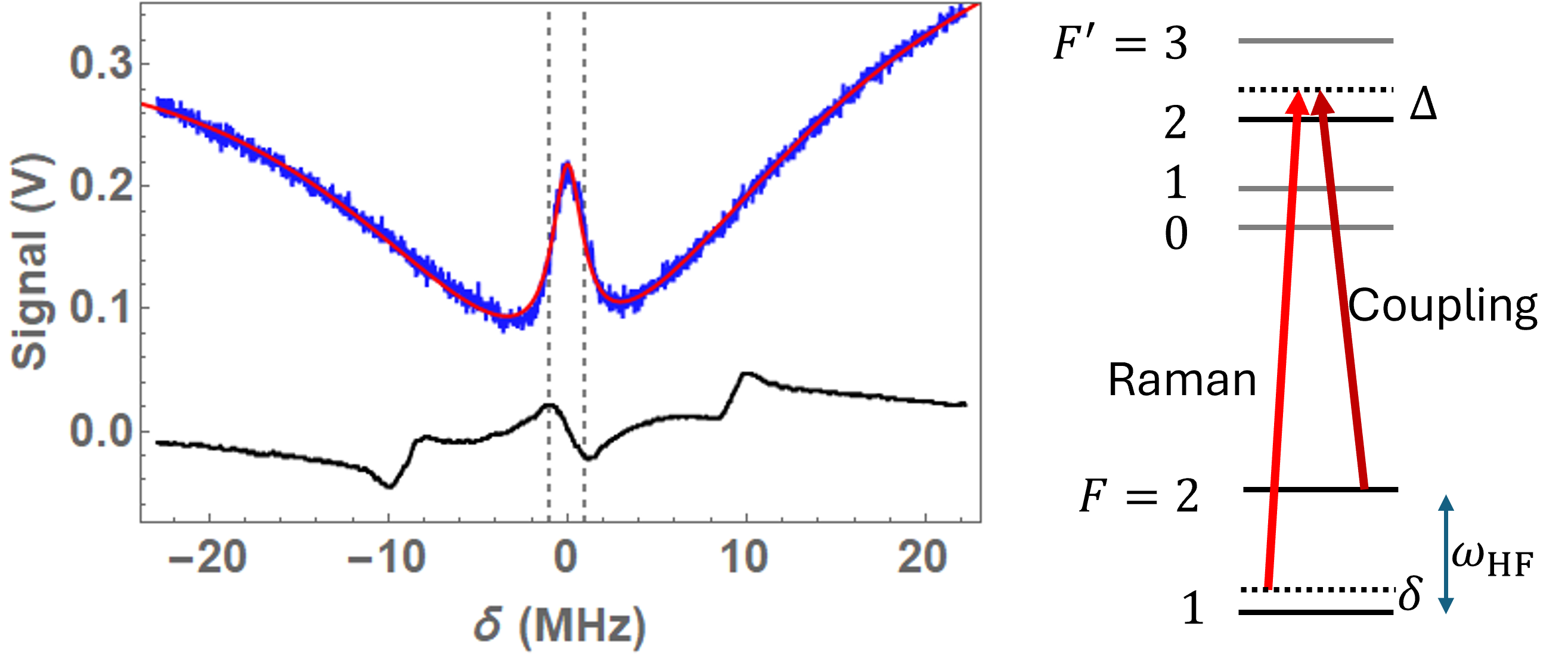}
    \caption{The EIT signal (blue) and the derived error signal (black), and the energy levels and laser scheme. The power-broadened width of the EIT feature signal is adjustable down to about 750 kHz, likely limited by the linewidth of the free running DL 100, which is nominally $\sim 1\,\text{MHz}$. In this figure, the EIT signal is fit with a sum of two Lorentzians and here gives an EIT-linewidth of $2.08\pm 0.03\,\text{ MHz}$. 
    The coupling laser is tuned blue of the $F=2\to F^\prime=2$ transition by $\Delta$. The Raman laser is tuned very near to Raman resonance. }
    \label{fig:signal_errsignal}
\end{figure}

 Upon exiting the cell, a second quarter-wave plate restores their orthogonal linear polarizations. The beams are subsequently separated by a beam splitter, and the probe beam is focused onto a fast, AC-coupled photodiode. The photodiode output is mixed with the $10\,\text{MHz}$ local oscillator signal and low-pass filtered to generate an error signal. Near the EIT resonance (see Figure~\ref{fig:signal_errsignal}), the error signal is proportional to the detuning, $\delta$, of the lasers from the Raman resonance condition. The width of the EIT signal, and hence the slope of the error signal, is determined by adjusting the probe and coupling laser power. The error signal is amplified by a TOPTICA FALC 110 (Fast Analog Linewidth Controller) , which provides flexible control over the feedback loop gain. The low-frequency component of the correction signal is sent to the piezoelectric actuator of the DL 100 to adjust the external cavity length, while the high-frequency component is fed back to the diode laser current via a high-bandwidth bias-T. After amplification we adjust laser powers to obtain 25 mV/MHz sensitivity. The 40 KHz modulation of the DL PRO laser does limit the width of the observed EIT feature. To obtain the best stability, we reduced the amplitude of the modulation as much as possible, which suggests the lock could be further improved by either electronically filtering the 40 KHz signal, or by locking the DL PRO to the saturated absorption signal using a scheme which avoids modulating the laser current.

\begin{figure}
    \centering
    \includegraphics[width=1.0\linewidth]{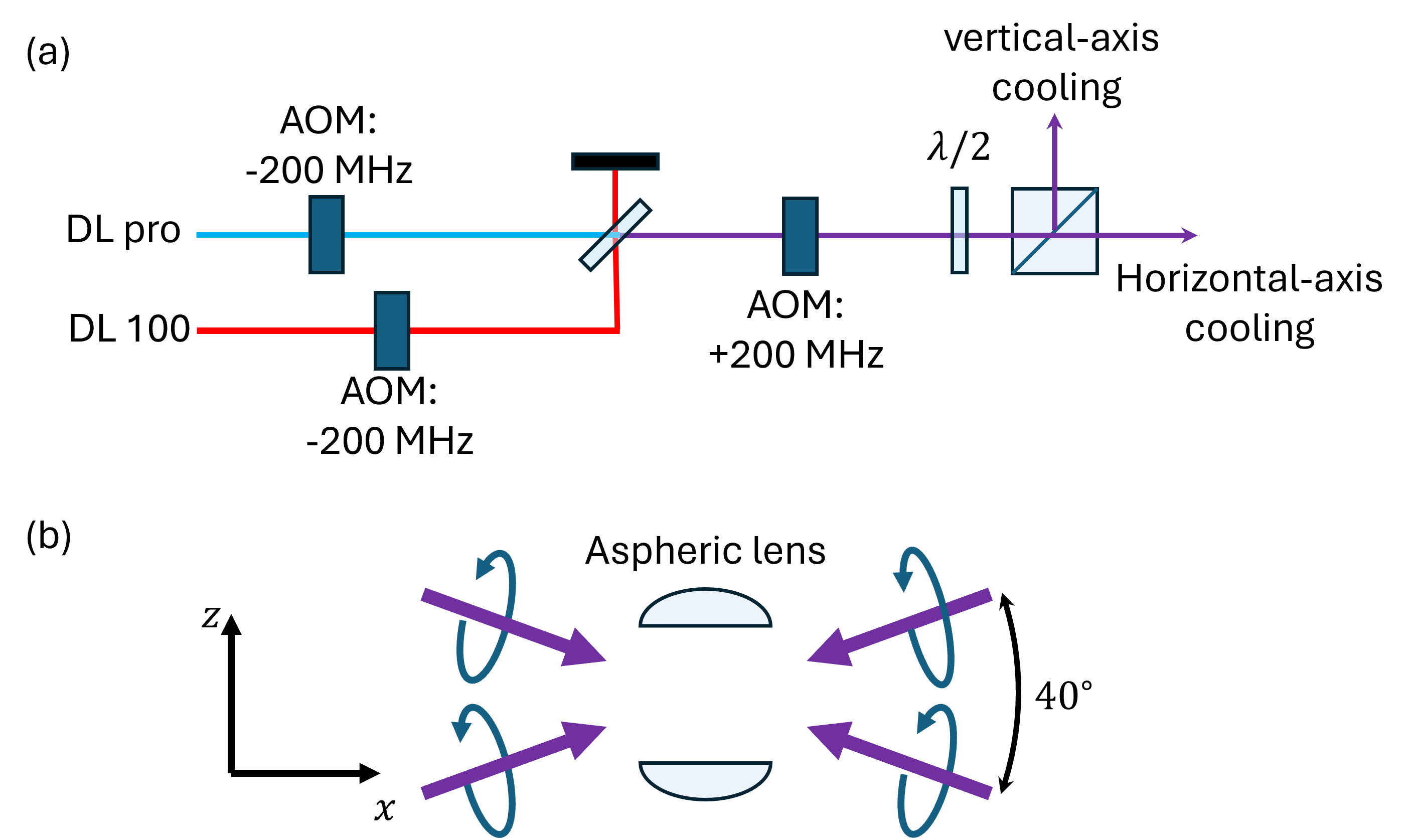}
    \caption{(a) The two gray molasses cooling beams are each down-shifted by $\sim 200\,\text{MHz}$ by an AOM, and then combined on a beam splitter. One of the output beams is selected for the experiment, and up-shifted by $\sim 160\,\text{MHz}$, so that the resulting cooling beams are shifted $\sim2\Gamma$ blue of the $F\to F^\prime = 2$ transitions. Finally, the beams are split on a polarizing beam splitter cube, with 1/3 of the incident power being sent to the vertical cooling beam and 2/3 sent to the horizontal cooling beams. In (b), the non-standard arrangement of the horizontal beams is depicted, showing the inclination of the beams from the horizontal $x-$axis and the location of the dipole beam shaping lenses symmetrically located on the $z$-axis. }
    \label{fig:beam_setup}
\end{figure}
\subsection{Pre-cooling in standard molasses}
The central aim of our experiment is to trap cold atoms in mesoscopic-scale dipole traps, which necessitates a laser cooling setup designed to accommodate a pair of high numerical aperture lenses. The lenses are mounted inside the vacuum chamber, restricting optical access for the four horizontal laser cooling beams. This constraint is common across many cold‑atom platforms used in quantum technologies. The beams have a waist of $\sim3.5 \,\text{mm}$ and the two counter-propagating pairs intersect at an angle of $40^\circ$, as shown in Figure \ref{fig:beam_setup}(b); additionally, the beams are weakly converging such that the intensity of the counter-propagating beams are balanced even after losses at the vacuum chamber windows. A repump beam tuned near the $F=1\to F^\prime=2$ transition is overlapped with the cooling beams, which are detuned 12 MHz from the $F=2\to F^\prime = 3$ transition. As a result, the small trapping volume of the Magneto optic trap leads to both small sample sizes (circa $6\times 10^5$ atoms), and our best efforts at Doppler and red-detuned sub Doppler have found temperatures no cooler than $T_\text{red}\sim 45\,\upmu\mathrm{K}$. 
\section{Gray molasses}
To improve on standard red-detuned optical molasses, we employed $\Lambda$-enhanced gray molasses, which is a technique that may allow sub-recoil temperatures (in Rb, $< 350 \,\mathrm{nK}$). 

The mechanism behind gray molasses is as follows: in the presence of two counter-propagating, orthogonally polarized laser beam pairs (driving two transitions between three states as depicted in Figure~\ref{fig:signal_errsignal}), an atom can be optically pumped into a dark state, i.e. a superposition of the two ground states and decoupled from the laser beams. When the lasers are blue detuned from a nearby transition, the dark state energy is lower than that of the bright states, whose energies are sinusoidally modulated with position. Motion of the atom in the potential energy landscape of the lasers leads to non-adiabatic passage of the atom from the dark state to the bright states, and the probability of such a transition peaks near the potential energy minima. An atom that transitions from the dark state at the potential minimum will climb the optical potential until it is optically pumped back to the dark state, losing the energy it gained climbing the potential hill.

In Rb$^{87}$, the hyperfine structure is more complicated than the three-level description given above, and the precise cooling mechanism is more complex. In this system, a strong \lq coupling\rq\, laser (with frequency $\omega_\text{C}$) is blue detuned from the $F=2\to F^\prime =2 $ cooling transition (which has frequency $\omega_{22}$) by $\Delta=\omega_{22}-\omega_\text{C}$,  and supplemented by a \lq Raman\rq\, laser (with frequency $\omega_\text{R}$), driving the $F=1\to F^\prime=2$ transition and tuned such that the two lasers are detuned from Raman resonance by $\delta=\omega_\text{HFS}+ (\omega_\text{R}-\omega_{C})$, where $\omega_\text{HFS} = 6.834 \,\text{GHz}$ is the frequency of the ground state hyperfine splitting (see Figure~\ref{fig:signal_errsignal}).  

Gray molasses is implemented by passing the DL Pro and DL 100 lasers through AOMs and then overlapping the beams with MOT lasers described above (see Figure~\ref{fig:beam_setup}). The AOMs allow control over both the absolute detuning and the Raman detuning, as well as the intensity of the two beams. As a result of our optical set up, both the coupling beam (with detuning $\Delta$ and peak intensity $I_0$) and the \lq Raman\rq\, beam (with detuning $\delta$ and peak intensity $I_{0,\text{R}}$) have the same $\sigma^-$ circular polarization. For the cooling beams, we choose $\Delta = 2\,\Gamma = 2\pi\times 12.1\,\text{MHz}$ and $I$ to obtain a maximum peak atom-light Rabi frequency, $\Omega_0= \Gamma\sqrt{I_0/2I_\text{sat}} = 3.5\,\Gamma%
$ (given the beam waist is $1.85\,\text{mm}$, each cooling beam has a power of $4.4\,\text{mW}$). For the Raman beams, we choose the beam power such that the peak rabi frequency $\omega_0 = \Gamma\sqrt{I_{0,\text{R}}/2I_\text{sat}}$ satisfies $\omega_0 / \Omega_0\approx \sqrt{0.1}$ (the Raman beam waist is $2.74\,\text{mm}$ and the power is typically $0.97\,\text{mW}$). In our initial experiments, we chose $\delta=0,$ but later we found $\delta=-0.1\,\Gamma$ was effective.
\section{Experimental results}
Pre-cooling in standard optical molasses prepares a 1.3 mm diameter cloud of $\sim 6\times 10^5\,\text{atoms}$ atoms at a temperature of $43\pm 2\,\upmu\mathrm{K}$. The atoms are then cooled in the gray molasses for 2--3 ms. After cooling, the gray molasses light is extinguished, and the atoms ballistically expand for a variable time $t$ until we fluorescence image the atoms with a 2 ms pulse of light resonant with the $F=2\to F^\prime=3$ transition. The fluorescence is captured using a 1:1 imaging system and an 8-bit CCD camera. As a result of our inefficient MOT, the atomic cloud has low density and therefore to achieve a desired signal to noise, we repeat the experiment $50\,-150$ times for each value of $t$. The cloud size $\sigma_i(t)$, $i=x,z$, is found by fitting a Gaussian profile to the images.
Our key metric is the cloud temperature, which we measure by a time of flight method. We do this by taking a series of images at various $t$, and fitting the measured cloud width (measured along the $z$-axis) with the relationship $\sigma_z^2(t) = \sigma_z(0)^2 + k_\mathrm{B}T t^2 /m$. 

\begin{figure}
    \centering
    \includegraphics[width=1.0\linewidth]{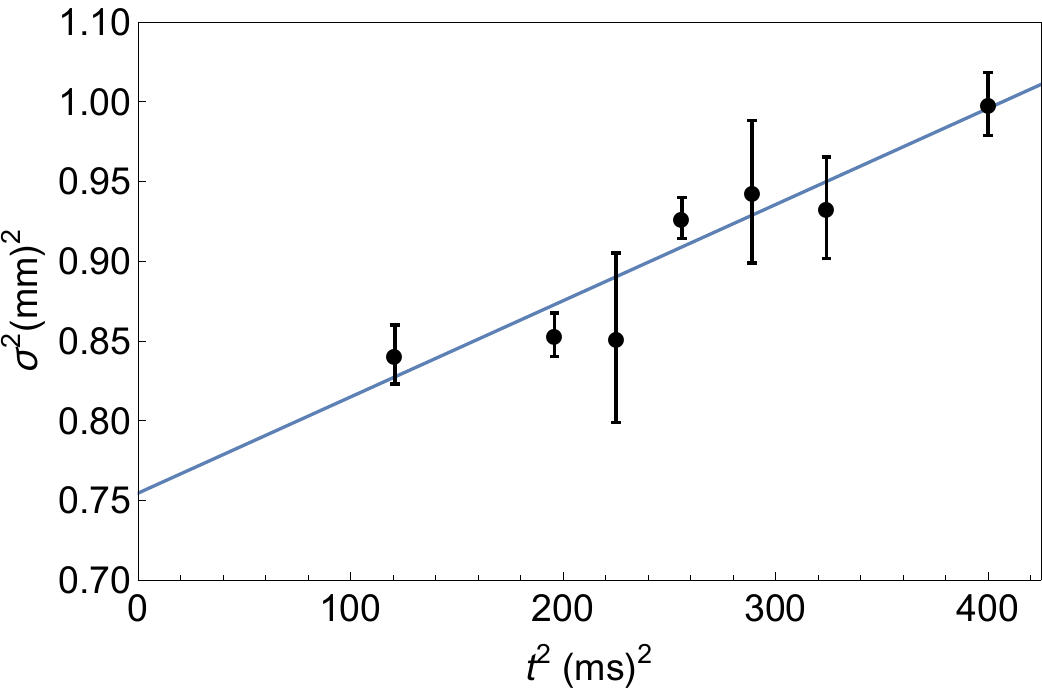}
    \caption{TOF analysis of the gray molasses cooled cloud for $\delta = -0.1 \Gamma$. The temperature, extracted from the least-squares best-fit, is $T=6.8\pm0.9\,\upmu\text{K}$, and represents an improvement by a factor of 1/7 over our standard optical molasses cooling. }
    \label{fig:main-result}
\end{figure}

Our initial experiments exploring the temperature dependence on $\Omega_0$, $\omega_0$ and $\delta$ were inconclusive because an instability in the sideband locking prevented the final temperature from reducing below $13\, \upmu\text{K}$, and hence hiding important features. Once stabilised, we saw temperatures as low as $\sim7\upmu\text{K}$, and we chose to fix $\Delta=2\Gamma$ but vary $\Omega_0$, $\omega_0$ and $\delta_\text{R}$. 
Once the sideband lock had been stabilised, we set $\Omega_0 = 2\Gamma$, $\omega_0= 0.27\,\Omega_0$ and $\delta=0$, we observed lower temperatures (below $10 \,\upmu\text{K}$). Following our Wave function Monte Carlo model(WFMC, see section \ref{sec-model}), we then chose $\delta = -0.1\,\Gamma$ and observed $T=6.8 \,\upmu\text{K}$ as shown in figure~\ref{fig:main-result}.

Since gray molasses is predicated on the interplay between the dark state and motional coupling to the bright states, we also expect to find that the population of the $\ket{F=2}$ ground states shows a dependency on the ratio $\omega/\Omega$ (here we drop the subscript `0' which defines the peak rabi frequency in a gaussian beam). To understand this dependency, we diagonalised the atom-light Hamiltonian (see Section \ref{sec-model}) over the $\ket{F_g=1,2,m_{F_g} }$ and $\ket{F_e = 2,m_{F_e} }$ states using standard techniques \cite{Sakurai1994}. The dark states $\psi_\text{D}$ are those ground states that (after diagonalization) do not project onto the excited $5P_{3/2}$ states. Since the dark states are in general a superposition of the $|F_g,m_F\rangle$ states, we can calculate the probability, $P_{F=2} = \sum_{m_F}\left\vert \langle F=2,m_F | \psi_\text{D} \rangle \right\vert^2$ of finding an atom in the $F=2$ states. In doing this, we found a family of dark states which can be expressed as a superposition of the ground state energy eigenfunctions, for example one state is \begin{equation}\label{eqn:F=2pop}
    \ket{\psi_\text{D} } =\frac1A\left[
    \ket{2,0}-\ket{2,\pm1}+\sqrt{\frac32}\ket{2,\pm2} \mp\frac{1}{\sqrt3\alpha}\ket{1,\pm 1} 
    \right]
\end{equation}
where $\alpha=\omega/\Omega$, and $A =\sqrt{6+2/3\alpha^2} $, giving an $F=2$ fractional population $P_{F=2} = 9\alpha^2/(1+9\alpha^2)$. We note that when including the $5P_{3/2}, F'=3$ states, there are no exact dark states because the Raman laser does not couple to $F'=3$, but the detuning of the cooling beam from the $F=3$ is large ($\approx 42\,\Gamma$) and so the dark state concept remains useful.   

\begin{figure}
    \centering
    \includegraphics[width=1.0\linewidth]{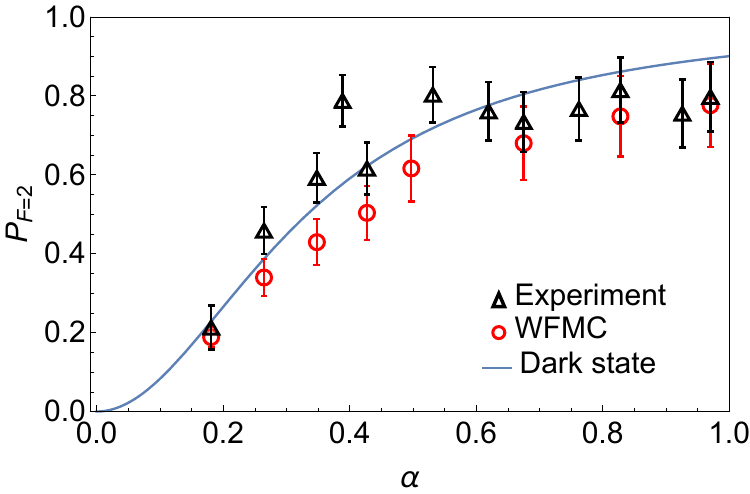}
    \caption{F=2 state population as a function of $\alpha=\omega/ \Omega$ for $\Omega = 2.0\, \Gamma$. The experimental points (black dots) are measured using state selective fluorescence. The continuous blue line is the fractional $5S_{3/2},F=2$ population obtained from Equation~(\ref{eqn:F=2pop}), $P_{F=2} = 9\alpha^2/(1+9\alpha^2)$, where $\alpha=\omega/\Omega$. The red points show the population calculated using the WFMC. Experimental uncertainties experimental data are estimated using a bootstrapping technique. The uncertainties in the WFMC method are $\Delta P = \langle P\rangle/\sqrt{N_\text{traj} } $ calculated over $N_\text{traj}=216$ trajectories.}
    \label{fig:F=2pop}
\end{figure}

In Figure~\ref{fig:F=2pop} the fractional population of the $\ket{F=2}$ state, after 2 ms of gray molasses, is plotted as a function of $\omega/\Omega$, together with the result obtained from a wave function Monte Carlo model of our experiment (described in Section IV) and the dark state population $P_{F=2}$. In our experiment, the cloud density is a gaussian, with size $\sigma \sim 0.900\,\text{mm}$ and the grey molasses beam waists are $w_\text{C} = 1.85\,\text{mm}$ for the cooling beam and $w_\text{R} = 2.74\,\text{mm}$ for the Raman beam. Therefore we define an average experimental $\Omega = 0.68 \,\Omega_0 $, where the factor of 0.68 scales the peak beam Rabi frequency to an average as experienced by the atoms. For the same reason, we define $\alpha=\omega/\Omega = 2.2\,\omega_0/\Omega_0$ .
Finally, we must account for the imperfection in our state-selective read out. 
When imaging the $5S_{1/2}F=2$ population alone, using the $5S_{1/2}, F=2 \to 5P_{3/2}, F'=3$ cycling transition for 1 ms, we lose some signal as the cloud is optically pumped into the $5S_{1/2}, F=1$ ground states. As a result, the measured signal is degraded by about 50-70\%. To confirm the relationship implied by Equation~\ref{eqn:F=2pop}, we fit the F=2 population to the line $P_{F=2} = \eta\, 9\alpha^2/(1+9\alpha^2)$, where $\eta$ is our imaging efficiency, which we find to be $\eta=0.55\pm0.02$. In figure~\ref{fig:F=2pop}
the agreement between the simulation and the experimental data provides further evidence for the underlying gray molasses dynamics, and supports conclusion that the dark state plays an important role in the cooling mechanism.
\section{Modelling}\label{sec-model}
The objective of our modelling is to understand how experimental imperfections limit the final temperature that we have achieved. There are many methods that have been used to model laser cooling mechanisms, notably the semi-classical force approach~\cite{Kosachev1994} and the wave function Monte-Carlo (WFMC) approach~\cite{Molmer1993}.

The dissipative effects of e.g. spontaneous emission can be modelled by solving the Lindblad Master equation for the system's density matrix $\rho$,
\begin{equation}
\label{eqn:Lindblad}
    \dot\rho = -i\left[ \operator{H},\rho\right] - \frac12 \sum_j  \left \{  \operator{L}^\dagger_j\operator{L}_j,\rho \right\}
    + \sum_j \operator{L}_j\rho\operator{L}^\dagger_j ,  
\end{equation}
where $\operator{H}$ is the Hamiltonian of the 24-level system (which drives coherent evolution of $\rho$) and $\operator{L}_j = \sqrt{\Gamma}\ket{g_j}\bra{ e_j}$ are jump operators which correspond to the possible spontaneous emission decay channels (labelled $j$). Additionally, we can include energy conserving dephasing e.g. due to laser noise, by including the additional terms $\operator{L}_j = \sqrt{\gamma}\ket{\psi_k}\bra{ \psi_k}$, with $\gamma$ being the laser-noise dephasing rate, and $\psi_j$ are ground/excited state level as appropriate to the dephasing process. 

However, to include a quantum description of the atomic motion increases the size of the Hilbert space to the point where the solving the Master equation is beyond our means. An alternative approach, the WFMC method~\cite{Molmer1993} is a method of approximately solving Equation~(\ref{eqn:Lindblad}). The advantage of his approach is that the time-evolution can be efficiently dealt with by exponentiating the Hamiltonian, which acts on the system wave function, and thus the computing resources (particularly memory) are reduced in comparison to numerically solving the Lindblad master equation. Thus the technique is valuable when considering the large Hilbert spaces spanned by $24$-level atoms with momenta spanning the range $\pm30 \hbar k$, quantised on a grid of $\hbar k/3$. 

The basis of the WFMC approach is the observation that the first two terms (i.e the commutator and the anti-commutator terms) of Equation~(\ref{eqn:Lindblad}) can be generated by the non-Hermitian Hamiltonian
 $   \operator{H}_\text{NH} = \operator{H} -\frac{i}{2} \sum_j  \operator{L}^\dagger_j \operator{L}_j ,
$
and hence this evolution of the wave-function can be found by applying the Schroedinger equation to the wave-function of the system.

The quantum jumps, however, cannot be accounted for in this manner. To handle the jumps (which are driven by the term $\sum_j \operator{L}_j \rho\operator{L}^\dagger_j $), they must be forced. First, the wave function is allowed to evolve according to
$
    \psi(t+\delta t) = \mathrm{exp}\left(-i\,\operator{H}_\text{NH} \delta t/\hbar\right) \,\psi(t).
$
Next, the probability that a quantum jump occurred during this interval, $p_e=\Gamma \delta t \sum_e |a_{e}|^2$ is compared to a randomly generated number $0<r<1$; if $p_e<r$, then the wave function $\psi(t+\delta t)$ is normalised and the process repeats for the next time step. Otherwise, $p_e>r$ and a quantum jump is forced by applying one of the operators $\operator{L}_j$, which must be chosen at random from an appropriate distribution. In the case of the full 24-level system, the appropriate distribution is found by computing $p_j = |L_j \psi(t+\delta t)|^2$, which gives the probability of each decay channel (calculated using $\psi(t+\delta t)$). Note that we must normalise $p_j$ such that $\sum_j p_j = 1$. Having picked a decay channel, the new wave function is calculated as $\psi(t+\delta t) \to \operator{L}_j \psi(t+\delta t) $, and then normalised. Finally, we must account for the momentum distribution of the scattered photons; projected on the to the $x$-axis at an angle $\theta_\text{sc}$, $\pi$-polarized photons scatter according to $P(\theta_\text{sc})\,\mathrm{d}\theta_\text{sc}=3\sin^2{\theta_\text{sc}}/8\pi$ and $\sigma$-polarized photons according to $P(\theta_\text{sc})\,\mathrm{d}\theta = 16\cos^2{\theta_\text{sc}}/2\pi$. Note that the polarization of the scattered photon is set when the decay channel is chosen, i.e. we keep track of the angular momentum change in the index $j$.

The Hamiltonian is given by
\begin{multline}
    \operator{H} = \hbar \sum_{i,\alpha} \frac{\omega_{il,\alpha}+\Omega_{il,\alpha}}{2}\ket{g_i,p}\bra{e_l,p+\alpha\hbar k} \\ + \sum_l (-\hbar\Delta+\varepsilon_{l}) \ket{e_l,p}\bra{e_l,p} \\ +\hbar\delta\sum_j \ket{g_j,p}\bra{g_j,p} +\sum_m \frac{p^2}{2M_\text{Rb}}\ket{\psi_m,p}\bra{\psi_m,p}+ h.c.,
\end{multline}
where $\alpha=\pm1$ indexes the beam direction propagation along $\pm z$, $i$ runs over the $5S_{1/2}$ $F=1,2$ ground states, $j$ over the $F=2$ ground states only, $l$ runs over the $5P_{3/2}$ excited states and $m$ runs over all 24 states; $\Omega_{il}$ is the Rabi frequency characterising the $F=2\to F^\prime =1,2,3$ atom--laser coupling and $\omega_{ik}$ is the Rabi frequency characterising the $F=1\to F^\prime =0,1,2$ atom--laser couplings; $\varepsilon_k$ is the hyperfine energy of the excited states with respect to the $F^\prime=2$ energy levels. For the non-Hermitian Hamiltonian, we write $\operator{H}_\text{NH} = \operator{H} -i\gamma/2\sum_{p,l}  \ket{e_l,p}\bra{e_l,p}$.

\begin{figure}
    \centering
    \includegraphics[width=1.0\linewidth]{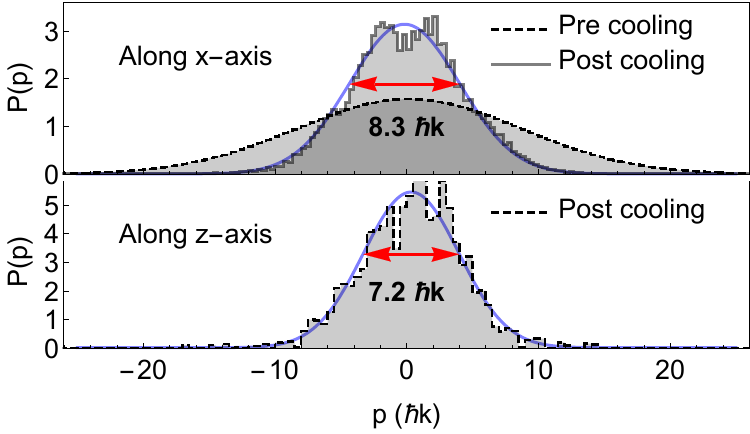}
    \caption{WFMC calculated final momentum distributions along the $x$ and $z$ directions using the parameters $\langle\Omega \rangle  =2.03\, \Gamma$, $\omega = \sqrt{0.1}\,\Omega$, $\Delta = 2.0\,\Gamma$ and $\delta = 0$. The initial distribution with $\langle \Delta p\rangle  = 12\,\hbar k $ cools to $\langle \Delta p_x\rangle  = 4.15\,\hbar k $ along the $x$ axis and $\langle \Delta p_z\rangle  = 3.60\,\hbar k $ along the $z$-axis after 1 ms of cooling. }
    \label{fig:WFMC-momdist}
\end{figure}

\vspace{0.2cm}
\emph{Modelling results and discussion}
In figure~\ref{fig:WFMC-momdist}, we show the momentum distributions calculated after cooling for $2\,\text{ms}$ with parameters $\Delta=2\,\Gamma$, $\Omega=1.5\,\Gamma = 3.16\,\omega $ and $\delta=0$. The Rabi frequency $\Omega=1.5\,\Gamma$ was chosen to reflect the fact that, as discussed above, most atoms in the experiment experience laser intensities lower than the peak intensity at the beam center. In these simulations, we assume perfect phase coherence between the lasers. The initial momentum distribution is a Gaussian with a standard deviation $\Delta p =12 \,\hbar k$, where $k=2\pi/780\,\text{nm}$ is the cooling laser wavelength.  Along the $z$-axis (see figure~\ref{fig:beam_setup}), the cooling is slightly stronger and the final momentum width is $\Delta p_z=3.6 \,\hbar k$. Along the $x$ axis, the final momentum reaches $\Delta p_x=4.15\,\hbar k$.
Since we can't model the 3D system, the 3D temperature is estimated from the 1D momentum distributions, taking $\Delta p_y=\Delta p_z$, yielding an estimated temperature of $T \approx 0.18(2\Delta p_x^2 + \Delta p_z^2) = 8.5\,\upmu\text{K}$, which is in good agreement with our experiments. Using the same parameters, we obtained $T\sim 10\,\upmu\text{K}$. 

\begin{figure}
    \centering
    \includegraphics[width=1.0\linewidth]{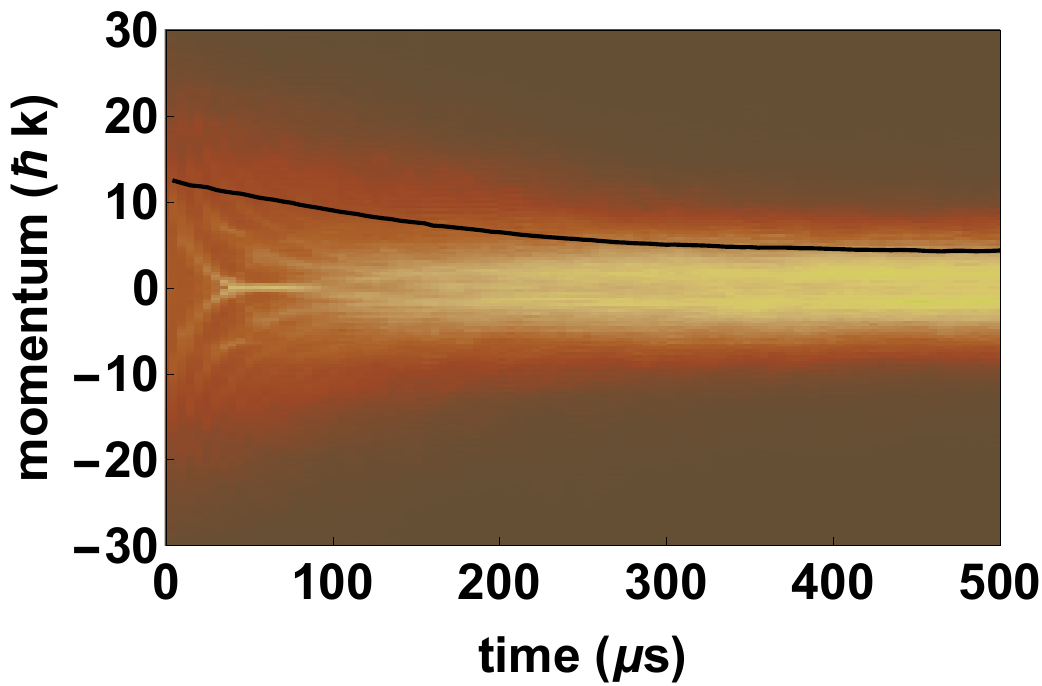}
    \caption{WFMC calculated cooling for $\langle\Omega \rangle  =2.03\, \Gamma$, $\omega = \sqrt{0.1}\,\Omega$, $\Delta = 2.0\,\Gamma$ and $\delta = 0$. The time evolution of momentum distribution with $\langle \Delta p\rangle  = 12\,\hbar k $ is shown. The black line indicates the $\langle \Delta p \rangle$ in units of $\hbar k$.}
    \label{fig:WFMC-evolution}
\end{figure}

In figure \ref{fig:WFMC-evolution} we show the rapidity of the cooling along the $x$-direction. The initial distribution, with width $\Delta p_{x}=12 \hbar k$ is cooled to $\sim4\hbar k$ after $500\,\upmu\text{s}$. We have found that the cooling along the $z$-axis is slower, taking around $1000\,\upmu\text{s}$ to reach $\sim3.6\,\hbar k$. 

We also considered experimental imperfections, namely external magnetic field strengths in the range $0\leq B_z \leq 4\times 10^{-5}\,\text{T}$ and a finite `linewidth' ($\Gamma_\text{R})$ of the Raman detuning, which mimics imperfection in the sideband lock. The magnetic field was included in the WFMC model by considering additional Zeeman terms $\Delta E_\text{B} = m_F g_F \mu_B B_z \, |5 L_J, F, m_F \rangle \langle 5L_J,F,m_F| $, where $m_F$ and $g_F$ are the magnetic quantum number the Land\'{e} g-factor respectively for the sub-level $|5L_J, F,m_F\rangle$,  and $B_z$ is the magnetic field strength for a magnetic field chosen to align with the $z-$axis. To account for finite Raman linewidth, we introduced discrete laser phase jumps at a rate $\Gamma_\text{R}$ into the WFMC quantum jump procedure. When the WFMC forced a quantum jump due to laser noise, the $F=1$ ground state manifold was given a random phase (chosen from a Gaussian distribution of width $\Gamma_\text{R}$) and the detuning $\delta$ appearing in the Hamiltonian was modified to the new value of $\delta$.

\begin{figure}
    \centering
    \includegraphics[width=1.0\linewidth]{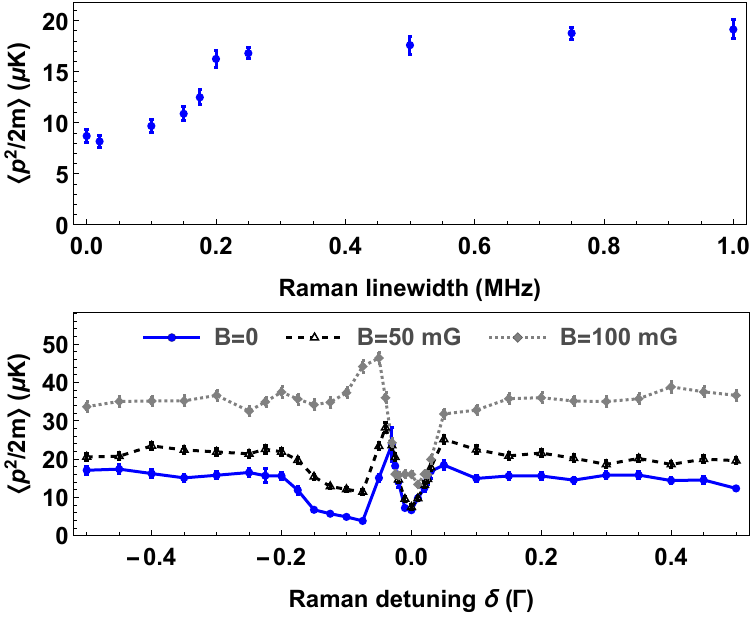}
    \caption{WFMC models with varying Raman linewidth $\Gamma_\text{R}$, magnetic field strength $B$ and Raman detuning $\delta$. In the upper panel, the final 3D temperature is plotted against the Raman laser linewidth. In the lower panel, the final 3D temperature is shown as a function of the Raman detuning $\delta$, and with $\Gamma_\text{R}=0$, each set of points representing WFMC data obtained for various values of $B$.}
    \label{fig:raman_inewidth}
\end{figure}

In the upper panel of Figure~\ref{fig:raman_inewidth}, we show the effects of finite Raman linewidth. Our results suggest that that as long as the lasers are locked to better than $\Gamma_\text{R}< 50 \,\text{kHz}$, then the final 3D temperature would be very similar to the that obtained using phase locked lasers. When $\Gamma_\text{R}\sim1\,\text{MHz}$, corresponding to the lasers being unlocked, the 3D temperature will approach $\sim 17 \,\upmu\text{K}$. This result agrees with the results of~\cite{rosi2018}, and is agreement with the temperatures we obtained when the sideband lock was unstable. Unfortunately, we don't at this time have a reliable way to measure $\Gamma_\text{R}$. However, our measurements give results comparable to those obtained for $\Gamma_R <50\, \text{KHz}$  demonstrating the efficiency of this implementation and the locking scheme.
Lastly, in the lower panel of Figure~\ref{fig:raman_inewidth}, we show the final temperature at varying $\delta$, at three different magnetic fields. Considering first the case of $B=0$, we note that along with the expected dip in $T$ near $\delta=0$ (where the system is Raman resonant, and hence there should be Lambda-enhanced cooling), we find a broader, low temperature dip between $\delta=-0.2 \,\Gamma$ and $\delta=-0.075 \,\Gamma$. For non-zero magnetic fields, the temperature increases but the Raman resonance at $\delta=0$  remains. The broad dip is washed out for $B=100\,\text{mG}$. Considering all the results shown in figure~\ref{fig:raman_inewidth}, we see that only for fields well below 50 mG, and for linewidths below 50 kHz, Lambda-enhanced cooling below $10\,\upmu\text{K}$ will be obtained. We see that the best cooling is observed when $\delta\sim -0.07\,\Gamma$, where we find $T=3.88\,\upmu\text{K}$. The model is not fully realistic, as it does not incorporate all experimental constraints. Therefore, a simplified model that neglects most experimental imperfections is not expected to exactly reproduce the experimental results. Good agreement with our experiment is found at $\delta=-0.1\,\Gamma$, where the model predicts $T=4.9\,\upmu\text{K}$ and we obtain  $6.8\pm0.9\,\upmu\text{K}.$ 


\section{Conclusion}
Our experiments show that Lambda enhanced Raman cooling is applicable to laser beam geometries which are sub-optimal for regular red-detuned laser cooling. The results of our experiments and modelling further indicate that our simplified laser lock, which requires no GHz electronics, is no barrier to obtaining very cold samples close to the limits of the Lambda-enhanced cooling technique. The relative simplicity and low cost of this technique will allow many labs to exploit lambda-enhanced gray molasses without costly or complicated new equipment. In the future, we hope to carry out a detailed study of the temperature dependence on the Raman detuning and investigate the broader resonance in Figure 8 centred near $\delta=-0.1\Gamma$.

\bibliography{expt_paper}
\end{document}